%% file: NIPS_workshop.tex
\documentclass{article}

\usepackage[final]{nips_2018}

\usepackage[utf8]{inputenc} % allow utf-8 input
\usepackage[T1]{fontenc}    % use 8-bit T1 fonts
\usepackage{hyperref}       % hyperlinks
\usepackage{url}            % simple URL typesetting
\usepackage{booktabs}       % professional-quality tables
\usepackage{amsfonts}       % blackboard math symbols
\usepackage{nicefrac}       % compact symbols for 1/2, etc.
\usepackage{microtype}      % microtypography
\input{math_command.tex}
\usepackage{graphicx}
\usepackage{natbib}
\usepackage{multicol}

\usepackage{xcolor}
\usepackage{enumitem}

\title{Correspondence Analysis of Government Expenditure Patterns}

\author{
Hsiang Hsu, Flavio P. Calmon, Jos\'e C\^andido Silveira Santos Filho\footnotemark\\
John A. Paulson School of Engineering and Applied Sciences\\
Harvard University\\
Cambridge, MA 02138\\
\texttt{\{hsianghsu, fcalmon, candido\}@g.harvard.edu}\\
\And
Andre P. Calmon\\
Technology and Operations Management\\
INSEAD\\
Fontainebleau, France\\
\texttt{andre.calmon@insead.edu}\\
\And
Salman Salamatian\\
Research Laboratory of Electronics\\
Massachusetts Institute of Technology\\
Cambridge, MA 02139\\
\texttt{salmansa@mit.edu}\\
}

\begin{document}
{
\renewcommand{\thefootnote}{*}
\footnotetext{J. C. S. Santos Filho is also with the Department of Communications, School of Electrical and Computer Engineering, University of Campinas, Campinas, SP, Brazil (e-mail: \texttt{candido@decom.fee.unicamp.br}).}
}
\maketitle

% \vspace{-.35in}
\begin{abstract}
%   \vspace{-.5em}
  We analyze expenditure patterns of discretionary funds by Brazilian congress members. This analysis is based on a large dataset containing over $7$ million expenses made publicly available by the Brazilian government. This dataset has, up to now, remained widely untouched by machine learning methods. Our main contributions are two-fold: (i) we provide a novel dataset benchmark for machine learning-based efforts for government transparency to the broader research community, and (ii) introduce a neural network-based approach for analyzing and visualizing outlying expense patterns. Our hope is that the approach presented here can inspire new machine learning methodologies for government transparency applicable to other developing nations.
\end{abstract}

% Outlines: 
% New datasets (made available after acceptance) - structure
% NN-based CA for visualizing expenditure - method and advantages vs PCA CCA
% describe potential use cases - predict corruption from spending pattern, watchdog efforts, correlation within parties, transfer to other countries.

% \vspace{-2em}
\section{Introduction}
% \vspace{-1em}
% public administration using machine learning Expenditure pattern
Over the last decade, an increasing number of the World's governments and, in particular, the executive and legislative branches of these governments, have made data on their activities and expenditures publicly available \citep{bates2012modern}.  This government-led \emph{open data} movement seeks to increase transparency, reduce corruption, make government activities more accessible to citizens and, ultimately, strengthen democratic institutions \citep{janssen2012benefits}. 

The open data trend in the public sector has led to many data science and machine learning (ML) based initiatives that seek to quantify, model, and evaluate the performance of public administration. In particular, \emph{Public Expenditure Analysis} (PEA) \citep{shah2005public}, which investigates how government budgets are spent, has become an active area of research in social and political science \citep{lopez2016performance, de2017longitudinal, garry2017analysis, odhiambo2018public}. 

Within this context, the goal of this paper is to apply machine learning  tools to perform PEA on data from a developing country whose executive and legislative branches have recently been marred by multiple budget misuse problems \citep{winter2017brazil, congressoemfoco}. 
Specifically, we apply a neural network(NN)-based technique \citep{hsu2018deep} to examine, visualize, and interpret the expenditure of discretionary funds by congress members of the Brazilian House of Congress (\emph{C\^amara dos Deputados}). 
This data has been made publicly available by the Brazilian government for about ten years, yet remains widely untouched by advanced ML techniques. 
We have translated the dataset to English and made it publicly available to the broader research community through the accompanying repository \citep{camarabrazil2018}. Our hope is that this dataset serves as a benchmark for new methodological ML-based approaches for ensuring and evaluating government transparency. We note that, more often than not, open government data is analyzed using a ``descriptive'' approach (e.g., finding outlier expenses, computing aggregate expenditure per congress member), as opposed to using more systematic ML techniques. Our ultimate goal is to reverse this trend. 

There are a few reasons why we focus on the discretionary expenses by the Brazilian Congress. First, the Brazilian government has a large open data initiative \citep{brazilopen}. Up to now, this data has been analyzed mostly through descriptive analytics. For example, the \emph{Opera\c{c}\~ao Serenata de Amor} \citep{deamor} has cleaned this data and made it available in a format that is easy to analyze, yet we are unaware of any efforts that use advanced ML techniques directly. 
Second, over the last few years many members of the Brazilian congress have been involved in high-profile budget misuse problems which have made global headlines \citep{winter2017brazil}, creating a natural test dataset for identifying budget misuse (some reports indicate that over $30\%$ of Brazilian congress members as of $2018$ are under investigation \citep{congressoemfoco, congressoemfoco2018}). This data can be used to validate methodological approaches that are adaptable to other countries.

%Finally, although Brazil is one of the largest democracies in the World, it is still a developing country, and this workshop provides a valuable opportunity for NIPS participants that are interested in or that originate from the developing world and want to investigate how data-science and machine learning tools can have positive social impact.

From a methodological standpoint, we use a generalization of   \emph{Correspondence Analysis} (CA) to continuous variables and high-dimensional data to visualize and interpret expenditure patterns by congress members. This approach is more suitable for the investigated dataset than traditional methods such as Principal Components Analysis (PCA) and Canonical Correlation Analysis \citep{hotelling1936relations}.
The potential use cases of the data and the method we present are four-fold:
\begin{enumerate}
    % \vspace{-.5em}
    \item Anomalous expenditure discovery and prediction in order to perform proactive reactions against budget misuse.
    % \vspace{-.5em}
    \item Clustering of congress members in terms of their discretionary expenditure pattern.
    % \vspace{-.5em}
    \item Interpretation and visualization of the expenditures, creation of algorithmic watchdogs for misuse.
    % \vspace{-.5em}
    \item Inspiration for new methodological approaches for government transparency transferable to other civic projects that aim at similar goals. 
    % \vspace{-.5em}
\end{enumerate}
All codes for downloading, translating and parsing the dataset is available at \citep{camarabrazil2018}; the dataset itself is made available by Brazilian government in \citep{deamor}. 
In the rest of this paper, we first describe the main ML tool used, namely CA using neural networks (Section~\ref{sec:nn}), and then describe the dataset and numerical results (Section~\ref{sec:exp}).
\section{Neural Network-based CA for Visualizing and Interpreting Expenditure}\label{sec:nn}
% \vspace{-1em}
CA is an exploratory multivariate statistical technique that converts data into a graphical display with orthogonal factors.
In a similar vein to PCA and its kernel variants \citep{hoffmann2007kernel}, CA is a technique that maps the data onto a low-dimensional representation. By construction, this new representation captures possibly non-linear relationships between the underlying variables, and can be used to interpret the dependence between two random variables $X$ and $Y$ from observed samples. CA has the  ability to produce interpretable,  low-dimensional visualizations (often two-dimensional) that capture complex relationships in data with entangled and intricate dependencies. This has led to its successful deployment in fields ranging from genealogy and epidemiology to social and environmental sciences \citep{tekaia2016genome, sourial2010correspondence, carrington2005models, ter2004co, ormoli2015diversity, ferrari2016whole}.

% \vspace{-.5em}
% \subsection{Correspondence Analysis}
% \vspace{-.5em}
CA considers two random variables $X$ and $Y$ with $|\calX| < \infty$, $|\calY| < \infty$, and their joint distribution $P_{X,Y}$ (cf. \citet{greenacre1984theory} for a detailed overview). Given samples $\{x_k, y_k\}_{k=1}^n$ drawn independently from $P_{X,Y}$, a two-way contingency table $\mathbf{P}_{X, Y}$ is defined as a matrix with $|\calX|$ rows and $|\calY|$ columns of normalized co-occurrence counts, i.e. $[\mathbf{P}_{X,Y}]_{i,j}=(\mbox{\# of observations } (x_i,y_i)=(i,j))/n$. 
Moreover, the marginals are defined as $\mathbf{p}_X \triangleq \mathbf{P}_{X, Y} \mathbf{1}_{|\mathcal{Y}|}$ and $\mathbf{p}_Y \triangleq \mathbf{P}_{X, Y}^T \mathbf{1}_{|\mathcal{X}|}$. 
Consider a matrix $\mathbf{Q}\triangleq \mathbf{D}_{X}^{-1/2}(\mathbf{P}_{X,Y}-\mathbf{p}_X\mathbf{p}_Y^T)\mathbf{D}_{Y}^{-1/2}$,
% \begin{equation}
% \label{eq:Q}
% \mathbf{Q}\triangleq \mathbf{D}_{X}^{-1/2}(\mathbf{P}_{X,Y}-\mathbf{p}_X\mathbf{p}_Y^T)\mathbf{D}_{Y}^{-1/2},
% \end{equation}
where $\mathbf{D}_{X} \triangleq \mathsf{diag}(\mathbf{p}_X)$ and $\mathbf{D}_{Y} \triangleq \mathsf{diag}(\mathbf{p}_Y)$, and let the singular value decomposition  of $\mathbf{Q}$ be $\mathbf{Q} = \bU \bSigma \bV^\intercal$. Let $d = \min\{ |\calX|, |\calY| \}-1$, and $\{\sigma_i\}_{i=1}^d$ be the singular values, then we have the following definitions \citep{greenacre1984theory}:
% \begin{multicols}{2}
\begin{itemize}
    % \vspace{-.5em}
    % \begin{minipage}{0.5\linewidth}
    \item Orthogonal factors of $X$: $\mathbf{L} \triangleq \mathbf{D}_{X}^{-1/2} \bU$.
    % \vspace{-.5em}
    \item Orthogonal factors of $Y$: $\mathbf{R} \triangleq \mathbf{D}_{Y}^{-1/2} \bV$.
    % \end{minipage}
    % \begin{minipage}{0.5\linewidth}
    \item Factor scores: $\lambda_i = \sigma_i^2, 1 \leq i \leq d$.
    % \vspace{-.5em}
    \item Factor score ratios: $\frac{\lambda_i}{\sum_{i=1}\lambda_i}, 1 \leq i \leq d$.
    % \end{minipage}
    % \vspace{-1.em}
\end{itemize}
% \end{multicols}
% CA makes use of the orthogonal factors $\mathbf{L}$ and $\mathbf{R}$  to visualize the correspondence (i.e., dependencies), between $X$ and $Y$. In particular, 
The first and second columns of $\mathbf{L}$ and $\mathbf{R}$ can be plotted on a two-dimensional plane (with each row corresponding to a point) producing the \emph{factoring plane}. The remaining planes can be produced by plotting the other columns of $\mathbf{L}$ and $\mathbf{R}$. The factor score ratio quantifies the correlations captured by each orthogonal factor, and is often shown along the axes in factoring planes. 

\textbf{Deep Neural Networks for Correspondence Analysis.} The contingency table-based approach for CA has three fundamental limitations.
First, it is restricted to data drawn from \emph{discrete} distributions with finite support, since  contingency tables for  continuous variables will be highly dependent on a chosen quantization which, in turn, may jeopardize information in the data.
Second, even when the underlying distribution of the data is discrete, reliably estimating the contingency table (i.e., approximating $P_{X,Y}$) may be infeasible due to limited number of samples. This inevitably hinges CA on the more statistically challenging problem of estimating $P_{X,Y}$.
Third, building contingency tables is not feasible for \emph{high-dimensional} data. This limitation can be circumvented by using a novel neural network-based approach for CA introduced in \citep{hsu2018deep}.

% Finding such functions, however, require a search over the space of all finite-variance functions of $X$ or $Y$, which is not feasible for high dimensional data. Thus, in order to approximate the principal functions and scale up CA, we restrict our search to \emph{functions representable by neural nets}. Note that the output of any neuron of a feed-forward neural net that receives $X$ as an input can be viewed\footnote{We assume that the outputs of a neural network have finite variance --- a reasonable assumption since several gates used in practice have bounded value (e.g., sigmoid, tanh) and, at the very least, the output is limited by the number of bits used in floating point representations.} as a point in $\calL_2(P_X)$ (and equivalently when receiving $Y$ as input). 

% In this section, we introduce the Correspondence Analysis Neural Net (CA-NN). The CA-NN estimates the PICs and the principal functions of $\Pxy$ by minimizing an appropriately defined loss function (described next) using gradient descent and backpropagation. We will use the CA-NN to perform CA at scale. 

Here, we summarize the neural network-based approach for CA in \citep{hsu2018deep}. 
Consider two neural networks F-Net and G-Net, which encode $X$ and $Y$ to $\Reals^d$ respectively. We denote the outputs from the F and G-Net of $X$ and $Y$, respectively, as
\begin{align}
    \mathbf{\tilde{f}}(X) \triangleq [\tilde{f}_1(X), \cdots, \tilde{f}_d(X)]^\intercal \in \Reals^{d\times 1}, \text{and}\; \mathbf{\tilde{g}}(Y) \triangleq [\tilde{g}_1(Y), \cdots, \tilde{g}_d(Y)]^\intercal \in \Reals^{d\times 1}.
\end{align}
The solution of the optimization problem
\begin{equation}\label{opti2}
\begin{aligned}
\min\limits_{\bA \in \Reals^{d\times d},\mathbf{\tilde{f}},\mathbf{\tilde{g}}} &\; \mathbb{E}\left[\|\bA\mathbf{\tilde{f}}(X)-\mathbf{\tilde{g}}(Y)\|^2_2\right],\quad \text{subject to}&\; \mathbb{E}\left[\bA\mathbf{\tilde{f}}(X)(\bA\mathbf{\tilde{f}}(X))^\intercal \right] = \mathbf{I}_d
\end{aligned}
\end{equation}
recovers the orthogonal factors of $X$ and $Y$ \citep{hsu2018deep}. 
%To see why this is the case, let $\bff(X)=\bA\mtf(X)=[\bff_1(X),\cdots,\bff_d(X)]^\intercal$ and suppose that $\bff,\mtg$ and $\bA$ achieve optimality in (\ref{opti2}). Optimality under quadratic loss implies that $\tilde{g}_i(y)=\EE{f_i(X)\mid Y=y}$ for $i\in \{1,\dots,d\}$. Moreover, the orthogonality constraint assures that the entries of $\bff(X)$ satisfy $\EE{f_i(X)f_j(X)}=\delta_{i,j}$, and thus form a basis for a  $d$-dimensional subspace of $\calL_2(\Px)$. As discussed in Section~\ref{sec:pic}, conditional expectation on $Y$ is a compact operator from $\calL_2(\Px)\to \calL_2(\Py)$, and from orthogonality of $\bff(X)$, it follows directly from the Hilbert-Schmidt Theorem \citep{reed1980functional} that the optimal value of (\ref{opti2}) is $\sum_{i=0}^{d-1}\lambda_i$, with $\bff$ and $\mtg$ corresponding to the $d$ largest principal functions.
Using theoretical results from orthogonal Procrustes problem \citep{gower2004procrustes}, we can further simplify the objective function (\ref{opti2}) into an unconstrained version: 
\begin{equation}\label{opti4}
\begin{aligned}
\min\limits_{\mathbf{\tilde{f}}, \mathbf{\tilde{g}}} && - 2\|\bC_f^{-\frac{1}{2}}\bC_{fg}\|_d + \mathbb{E}[\|\mathbf{\tilde{g}}(Y)\|^2_2],
\end{aligned}
\end{equation}
where $\bC_f = \mathbb{E}[ \mathbf{\tilde{f}}(X)\mathbf{\tilde{f}}(X)^\intercal ]$, $\bC_{fg} = \mathbb{E}[ \mathbf{\tilde{f}}(X)\mathbf{\tilde{g}}(Y)^\intercal ]$, and $\|\bZ\|_d$ is the $d$-th Ky-Fan norm, defined as the sum of the singular values of $\bZ$ \citep{horn1990matrix}. Denoting by $\bA$ and $\bB$ the whitening matrices for $\mathbf{\tilde{f}}(\bX)$ and $\mathbf{\tilde{g}}(\bY)$, the orthogonal factors of $X$ and $Y$ are given by $\bff(X) = [f_0(X), \cdots, f_d(X)]^\intercal = \bA\mtf(X)$ and $\bg(Y) = [g_0(Y), \cdots, g_d(Y)]^\intercal = \bB\mtg(Y)$. The factor score $\lambda_i$ is given by $\EE{f_i(X)^Tg_i(Y)}$, $1 \leq i \leq d$. 
The loss (\ref{opti4}) is unconstrained over the space of all finite variance functions of $X$ and $Y$, and therefore is trainable via back-propagation using the common loss function (\ref{opti4}).
% We restrict our search to functions in Hilbert spaces given by outputs of neural nets, parameterizing $\mathbf{\tilde{f}}(X)$ and $\mathbf{\tilde{g}}(Y)$ by  $\theta_F$ and $\theta_G$, respectively. Here, $\theta_F$ and $\theta_G$ denote weights of two neural nets, called the F-net and the G-net. The F-Net and the G-Net encode $X$ and $Y$ to $\Reals^d$, respectively, and are back-propagated using a common loss function (\ref{opti4}).
For more information about optimization details, see \citep{hsu2018deep}.

% \begin{figure*}[t!]
% \centering
% \includegraphics[width=.9\textwidth]{figures/PICE_net.pdf}
% \caption{{\small The architecture of the CA-NN, consisting of two encoders F-Net and G-Net for $X$ and $Y$ respectively to estimate the principal functions. The PIC loss is given by (\ref{opti4}).}}
% \label{fig:fg_nets}
% \end{figure*}

% I also developed mathematical tools based on Hilbert Spaces and, more broadly, functional analysis, to characterize representations learned by complex learning models. This Hilbert space approach uses an information-theoretic tool called the "principal inertia components", and can be used to guide the training of deep learning models in order to produce interpretable, large-scale data visualizations that capture the dependency between different feature in data.
% \vspace{-1em}
\section{The Data: Brazilian House of Congress Discretionary Spending}\label{sec:exp}
% \vspace{-1em}
\textbf{Description and Pre-processing of the Dataset.} We investigate data on discretionary funding reimbursements from the Brazilian House of Congress. This data was made openly and freely available (in Portuguese) by \cite{brazilopen}. Each Brazilian congress member receives a certain amount of discretionary funding for supporting parliamentary activity (\emph{Cota para o Exercício da Atividade Parlamentar -- CEAP)}~\citep{ceap}. This fund is used to reimburse travel, food, phone bills, postal services, cabinet costs, etc.  The limit that each congress member can spend depends on their state of origin, with a maximum monthly cap of around BRL\$$45$k (about USD\$$13$k) \citep{ceap}. Brazilian Congress has $513$ seats distributed among $26$ states and the Federal District. Brazil has several political parties, with over $30$ parties being represented in Congress as of $2018$. The term for a congress member is $4$ years. 

\begin{figure*}[t!]
\centering
\includegraphics[width=.99\textwidth]{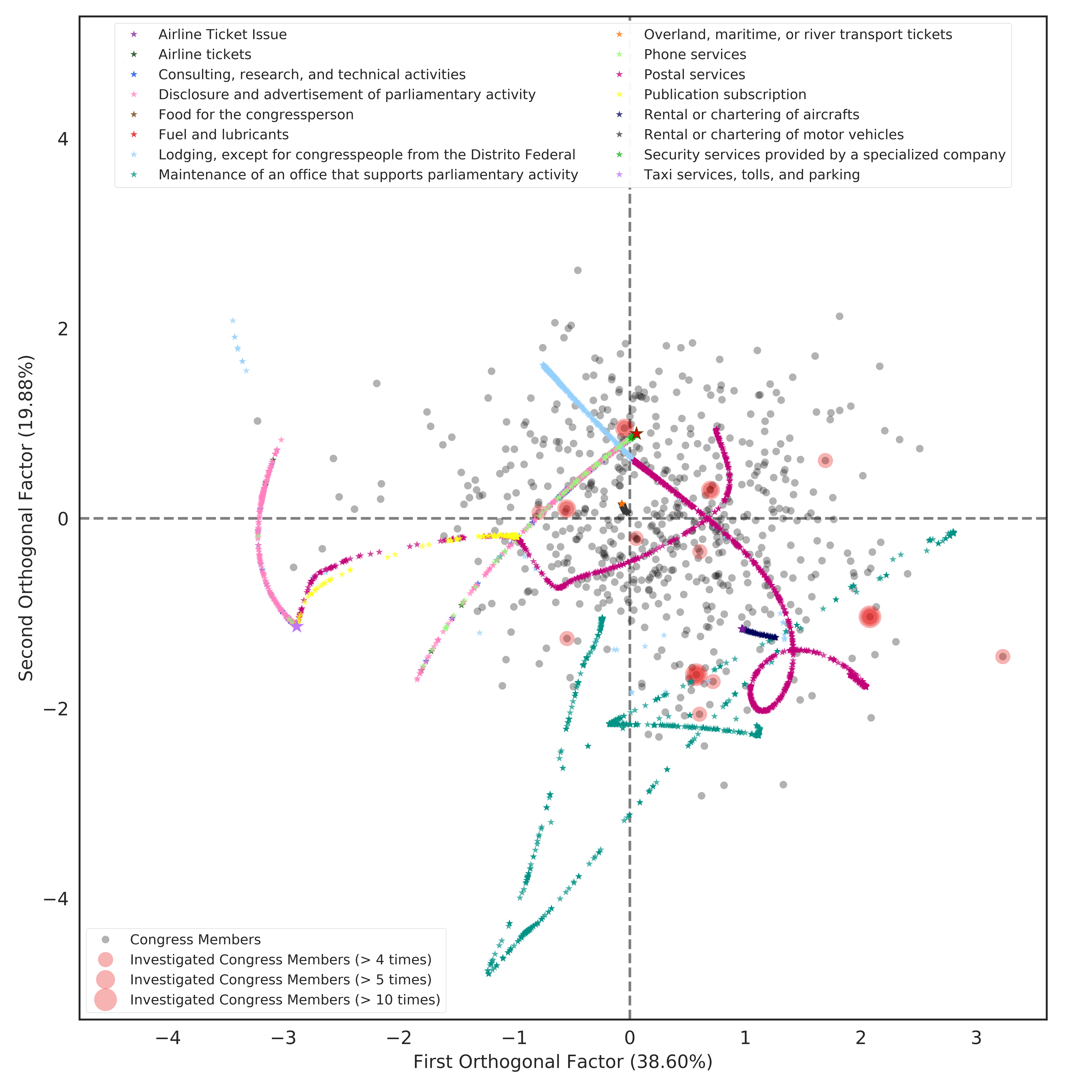}
\caption{{\small The first factoring plane of the expenditures of $595$ congress members in Brazil from $2015$ to $2018$. The factor score ratios are shown with the axis; higher factor score ratio means more correlation captured by the orthogonal factor. 
Each trace (e.g. pink line) represents the expenditure pattern of a category for all congress members. Each grey dot represents a congress member without investigations, and each red dot represents one under investigations according to \citep{congressoemfoco}, where the radius of red dots indicates the number of investigations (We did not independently verify the completeness/ correctness of the dataset, and recommend caution when using information about ongoing investigations to avoid potential errors (false positive) in analysis).
Points and lines close to the center (the origin) indicate small correlation. 
Names, states, and parties of congress members are omitted.}}
\label{fig:BC_CA}
\end{figure*}

We have produced code in Python for automatically downloading, translating and parsing this data, as well as meta-data regarding the multiple features found in the dataset, available at \citep{camarabrazil2018}.
The dataset contains more than $7$ million expenditure records from $2009$ to $2018$, including the category (e.g., fuel, food, office maintenance, airline tickets), values, date, and vendor that produced the receipt for the expenditure. Moreover, the states, parties, and names of the congress members are also included. In the analysis here, we present the records for the most recent term (i.e., $2015$--$2018$), dropped missing data points, and eliminated categories that appear less than $500$ times. The resulting dataset finally contains approximately $1.1$ million expenditure records in $16$ categories of $595$ congress members from  $26$ parties and $26$ states and the Federal District (the number of congress members is greater than the number of seats since not all members finish their term). For the CA, we set $X$ to be the categories and values of the expenditure and $Y$ to be the congress member with their parties and states, and perform a $70\%$-$30\%$ training-validation split of the data.

\textbf{Neural Network and Training Configuration.} The F and G-Net are composed of two simple feed-forward neural networks with different structures.
The F-Net has four layers with number of units $1000, 500, 300, 50$ and G-Net has three layers with number of units $100, 50, 30$.
We adopt \texttt{tanh} activation for hidden layers and the readout layer.
We train for $20$ epochs on the training set with a batch size of $256$ using a gradient descent optimizer with a learning rate of $0.01$. 
The result of the CA for expenditure analysis is shown in Fig.~\ref{fig:BC_CA}. 

\textbf{Expenditure Pattern.} In Fig.~\ref{fig:BC_CA} (see caption for instructions), we show the expenditure patterns of $16$ categories and the congress member in a standard CA factor-plane plot \citep{greenacre1984theory}. CA is performed using the NN-based approach described in the previous section. We summarize our observations below:
\begin{itemize}[leftmargin=*]
    % \vspace{-.5em}
    \item Our generalized CA approach automatically clusters related expenses together since they have close patterns, e.g., aviation-related expenses ``Airline Ticket Issue'' and ``Rental of aircrafts'', transportation-related expenses ``River transport tickets'' and ``Rental of motor vehicles'', and daily expenses ``Food'', ``Fuel'', and ``Security services''.
    \item There are certain categories of expenditures that are not correlated with specific congress members: ``Food'', ``Fuel and lubricants'', ``River transport tickets'', ``Rental of motor vehicles'', ``Security services'', and ``Taxi services and parking''. 
    % This indicates that most congress members behave similarly on these types of expenditures.
    \item Categories that show high variations also have clear pattern. For instance, overlapping traces of ``Publication subscription'' and ``Postal service'', and ``Airline tickets'', ``Consulting, research, and technical activities'' and ``Disclosure and advertisement of parliamentary activity'' can be observed. Moreover, ``Disclosure and advertisement of parliamentary activity'' has a very large variation (pink line on the left-hand side of the graph). This may potentially indicate mishandling of expenses in these categories.
    \item Two categories exhibit outlying patterns: ``Maintenance of an office'' and ``Lodging''. This might indicate that in different states, the expense on the two categories is dramatically different, or could be an indication of foul play. This can help direct further investigatory efforts.
    % \vspace{-.5em}
\end{itemize}

% \red{[this paragraph makes no sense]} The expenditure patterns in Fig.~\ref{fig:BC_CA} suggest even in different states and with different parties, in some categories, congresspeople spend their budgets with a common pattern. Therefore, anomaly expenditures are worth of extra attention or investigation. 

\textbf{Charged Congress members.} We also collected information from publicly available sources on congress members that are currently under investigation \citep{congressoemfoco, congressoemfoco2018}\footnote{We did not independently verify the completeness/ accuracy of the dataset, and recommend caution when using information about ongoing investigations to avoid potential errors (false positive) in analysis.}.
We display in Fig.~\ref{fig:BC_CA} those who are under multiple investigations. 
As we can see, the investigated congress members are concentrated near expenditure patterns that have large variation, i.e. outliers. This may indicate that congress members under multiple investigations also deviate from the mean use of discretionary funding, suggesting that discretionary funding may be predictive of misbehaviours  --- even though further investigation is required to confirm this statement. This analysis demonstrate how modern ML techniques can be applied to this large dataset to both visualize and interpret congress member behaviours.

% \section{Conclusion and Future Works}
% We propose a novel public expenditure analysis via neural network-based correspondence analysis, which allows us to visualize and interpret expenditure patterns of Brazilian House of Congress discretionary spending. 
% \textbf{Predicting corruption from spending pattern.}

% \textbf{Watchdog efforts.}

% \textbf{Correlation within parties.}

% \textbf{Transferring to other countries.}

% \clearpage
\bibliographystyle{apalike}
\bibliography{aistats2019.bib}
\end{document}

%% file: math_command.tex
\usepackage{amsmath,amsfonts,bm}
\usepackage{amssymb}

\def\eqref#1{equation~\ref{#1}}
\def\1{\bm{1}}

\DeclareMathAlphabet{\mathsfit}{\encodingdefault}{\sfdefault}{m}{sl}
\SetMathAlphabet{\mathsfit}{bold}{\encodingdefault}{\sfdefault}{bx}{n}

\newcommand{\EE}[1]{\mathbb{E}\left[#1\right]}

\newcommand{\calX}{\mathcal{X}}

\newcommand{\calY}{\mathcal{Y}}

\newcommand{\bff}{\mathbf{f}}
\newcommand{\bg}{\mathbf{g}}
\newcommand{\bSigma}{\bm{\Sigma}}

\newcommand{\bA}{\mathbf{A}}
\newcommand{\bB}{\mathbf{B}}
\newcommand{\bC}{\mathbf{C}}

\newcommand{\bU}{\mathbf{U}}
\newcommand{\bV}{\mathbf{V}}

\newcommand{\bY}{\mathbf{Y}}
\newcommand{\bZ}{\mathbf{Z}}

\newcommand{\bX}{\mathbf{X}}

\renewcommand{\tilde}{\widetilde}

\newcommand{\Reals}{\mathbb{R}}
\newcommand{\mtf}{\mathbf{\tilde{f}}}
\newcommand{\mtg}{\mathbf{\tilde{g}}}

\usepackage{algorithmicx}
\usepackage{algorithm}
\usepackage[noend]{algpseudocode}
\algnewcommand\algorithmicinput{\textbf{Input:}}
\algnewcommand\Input{\item[\algorithmicinput]}
\algnewcommand\algorithmicoutput{\textbf{Output:}}
\algnewcommand\Output{\item[\algorithmicoutput]}

\usepackage{amsthm}

\theoremstyle{remark}

%% file: NIPS_workshop.bbl
\begin{thebibliography}{}

\bibitem[Bates, 2012]{bates2012modern}
Bates, J. (2012).
\newblock “this is what modern deregulation looks like”: co-optation and
  contestation in the shaping of the uk’s open government data initiative.
\newblock {\em The Journal of Community Informatics}, 8(2).

\bibitem[Cagni, 2017]{congressoemfoco}
Cagni, P. (2017).
\newblock Os deputados sob investigação no supremo tribunal federal.
\newblock Congresso em Foco,
  \url{https://congressoemfoco.uol.com.br/especial/noticias/os-deputados-sob-investigacao-no-supremo-tribunal-federal/}.

\bibitem[Carrington et~al., 2005]{carrington2005models}
Carrington, P.~J., Scott, J., and Wasserman, S. (2005).
\newblock {\em Models and methods in social network analysis}, volume~28.
\newblock Cambridge university press.

\bibitem[de~Sousa et~al., 2017]{de2017longitudinal}
de~Sousa, R.~G., Paulo, E., and Mar{\^o}co, J. (2017).
\newblock Longitudinal factor analysis of public expenditure composition and
  human development in brazil after the 1988 constitution.
\newblock {\em Social Indicators Research}, 134(3):1009--1026.

\bibitem[Ferrari et~al., 2016]{ferrari2016whole}
Ferrari, A., Vincent-Salomon, A., Pivot, X., Sertier, A.-S., Thomas, E., Tonon,
  L., Boyault, S., Mulugeta, E., Treilleux, I., Macgrogan, G., et~al. (2016).
\newblock A whole-genome sequence and transcriptome perspective on
  her2-positive breast cancers.
\newblock {\em Nature communications}, 7:12222.

\bibitem[Garry and Rivas~Valdivia, 2017]{garry2017analysis}
Garry, S. and Rivas~Valdivia, J.~C. (2017).
\newblock An analysis of the contribution of public expenditure to economic
  growth and fiscal multipliers in mexico, central america and the dominican
  republic, 1990-2015.

\bibitem[Gower and Dijksterhuis, 2004]{gower2004procrustes}
Gower, J.~C. and Dijksterhuis, G.~B. (2004).
\newblock {\em Procrustes problems}, volume~30.
\newblock Oxford University Press on Demand.

\bibitem[Greenacre, 1984]{greenacre1984theory}
Greenacre, M.~J. (1984).
\newblock {\em Theory and applications of correspondence analysis}.
\newblock London (UK) Academic Press.

\bibitem[Hoffmann, 2007]{hoffmann2007kernel}
Hoffmann, H. (2007).
\newblock Kernel pca for novelty detection.
\newblock {\em Pattern recognition}, 40(3):863--874.

\bibitem[Horn et~al., 1990]{horn1990matrix}
Horn, R.~A., Horn, R.~A., and Johnson, C.~R. (1990).
\newblock {\em Matrix analysis}.
\newblock Cambridge university press.

\bibitem[Hotelling, 1936]{hotelling1936relations}
Hotelling, H. (1936).
\newblock Relations between two sets of variates.
\newblock {\em Biometrika}, 28(3/4):321--377.

\bibitem[Hsu and Calmon, 2018]{camarabrazil2018}
Hsu, H. and Calmon, F.~P. (2018).
\newblock Camara brazil.
\newblock \url{https://github.com/HsiangHsu/Brazilian-Congress-Expenditure}.

\bibitem[Hsu et~al., 2018]{hsu2018deep}
Hsu, H., Salamatian, S., and Calmon, F.~P. (2018).
\newblock Deep orthogonal representations: Fundamental properties and
  applications.
\newblock {\em arXiv preprint arXiv:1806.08449}.

\bibitem[Janssen et~al., 2012]{janssen2012benefits}
Janssen, M., Charalabidis, Y., and Zuiderwijk, A. (2012).
\newblock Benefits, adoption barriers and myths of open data and open
  government.
\newblock {\em Information systems management}, 29(4):258--268.

\bibitem[Lopez et~al., 2016]{lopez2016performance}
Lopez, G. H.~N., Mori, E.~S., Avila, L., Lozano, R., et~al. (2016).
\newblock A performance analysis of public expenditure on maternal health in
  mexico.

\bibitem[Musskopf, 2016]{deamor}
Musskopf, I. (2016).
\newblock Operation serenata de amor.
\newblock \url{https://serenata.ai/en/}.

\bibitem[Odhiambo, 2018]{odhiambo2018public}
Odhiambo, N.~M. (2018).
\newblock Public expenditure and economic growth in kenya: A multivariate
  dynamic causal linkage.

\bibitem[Ormoli et~al., 2015]{ormoli2015diversity}
Ormoli, L., Costa, C., Negri, S., Perenzin, M., and Vaccino, P. (2015).
\newblock Diversity trends in bread wheat in italy during the 20th century
  assessed by traditional and multivariate approaches.
\newblock {\em Scientific reports}, 5:8574.

\bibitem[Sardinha, 2018]{congressoemfoco2018}
Sardinha, E. (2018).
\newblock Um em cada três deputados é acusado de crimes.
\newblock Congresso em Foco,
  \url{https://congressoemfoco.uol.com.br/especial/noticias/um-em-cada-tres-deputados-e-acusado-de-crimes/}.

\bibitem[Shah, 2005]{shah2005public}
Shah, A. (2005).
\newblock {\em Public Expenditure Analysis}.
\newblock The World Bank.

\bibitem[Sourial et~al., 2010]{sourial2010correspondence}
Sourial, N., Wolfson, C., Zhu, B., Quail, J., Fletcher, J., Karunananthan, S.,
  Bandeen-Roche, K., B{\'e}land, F., and Bergman, H. (2010).
\newblock Correspondence analysis is a useful tool to uncover the relationships
  among categorical variables.
\newblock {\em Journal of clinical epidemiology}, 63(6):638--646.

\bibitem[Tekaia, 2016]{tekaia2016genome}
Tekaia, F. (2016).
\newblock Genome data exploration using correspondence analysis.
\newblock {\em Bioinformatics and Biology insights}, 10:BBI--S39614.

\bibitem[ter Braak and Schaffers, 2004]{ter2004co}
ter Braak, C.~J. and Schaffers, A.~P. (2004).
\newblock Co-correspondence analysis: a new ordination method to relate two
  community compositions.
\newblock {\em Ecology}, 85(3):834--846.

\bibitem[{The Brazilian House of Congress}, 2018]{ceap}
{The Brazilian House of Congress} (2018).
\newblock {Cota para o Exercício da Atividade Parlamentar -- CEAP}.
\newblock
  \url{http://www2.camara.leg.br/transparencia/acesso-a-informacao/copy\_of\_perguntas-frequentes/cota-para-o-exercicio-da-atividade-parlamentar}.

\bibitem[{The Brazilian Ministry of Planning}, 2012]{brazilopen}
{The Brazilian Ministry of Planning} (2012).
\newblock Portal brasileiro de dados abertos.
\newblock \url{http://dados.gov.br}.

\bibitem[Winter, 2017]{winter2017brazil}
Winter, B. (2017).
\newblock Brazil's never-ending corruption crisis: Why radical transparency is
  the only fix.
\newblock {\em Foreign Aff.}, 96:87.

\end{thebibliography}
